\documentclass[pra, twocolumn, amsfonts]{revtex4}
\usepackage{amsfonts}
\usepackage{amsmath}
\usepackage{bm}
\usepackage{graphicx}
\usepackage{epsf}
\usepackage{amssymb}
\usepackage{dsfont}

\begin{document}

\title{An Arrow of Time Operator for Standard Quantum Mechanics}

\author{Y. Strauss$^{1}$,
J. Silman$^{2}$, S. Machnes$^{2}$, and L.P. Horwitz$^{2, 3, 4}$}
\affiliation{$^{1}$Einstein Institute of Mathematics, Edmond J. Safra campus,
The Hebrew University of Jerusalem, Jerusalem 91904, Israel
\\
 $^{2}$ School of Physics and Astronomy, Raymond and Beverly Sackler
Faculty of Exact Sciences,
Tel-Aviv University, Tel-Aviv 69978, Israel\\
 $^{3}$ Physics Department of Physics, Bar-Ilan University, Ramat-Gan 52900, Israel\\
 $^{4}$Department of Physics, The Ariel University Center of Samaria, Ariel 40700, Israel}
\date{February 17, 2008}

\begin{abstract}
\textbf{We introduce a self-adjoint operator that indicates the direction of time within the framework of standard quantum mechanics.
That is, as a function of time its expectation value decreases monotonically for any initial state. This operator can be defined for any system governed by a Hamiltonian with a uniformly finitely degenerate, absolutely continuous and semibounded spectrum. We study some of the operator's properties and illustrate them for a large equivalence class of scattering problems. We also discuss some previous attempts to construct such an operator, and show 
that the no-go theorems developed in this context are not applicable to our construction.}
\end{abstract}
\maketitle

The emergence of irreversible phenomena in systems governed by reversible dynamical laws is a fundamental problem for both classical and quantum physics. In this context, the construction of a quantity indicating the direction of time is one of the central goals. As a functional of the state of the system, such a quantity must vary monotonically in time.  A quantity having this property is often termed a Lyapunov functional. The need for a Lyapunov functional arises in different quantum mechanical problems such as the decay of a metastable state, resonant processes, and other  irreversible phenomena. The main result of this paper is the explicit construction of a Lyapunov operator -- an operator whose expectation value decreases monotonically independently of the initial state -- within the framework of standard quantum mechanics.\\

The question of whether a Lyapunov functional can be defined for classical and quantum Hamiltonian dynamical systems on phase space and  Hilbert space, respectively, has been investigated by various authors. An early and fundamental theorem by Poincar\'{e} states that no "local"  function on phase space can give rise to a quantity with the characteristics of nonequilibrium entropy, i.e. a Lyapunov functional \cite{Poincare}. Close to a century later, in the late 70's, it was demonstrated by Misra that if this restriction on the phase space functions is lifted, a Lyapunov functional can be constructed \cite{Misra0}. Soon after, in an attempt to generalize this result, Misra, Prigogine, and Courbage (MPC) published a paper containing a proof which was taken by many to imply that standard quantum mechanics does not allow for a Lyapunov operator \cite{Misra}. This proof is based on a number of assumptions, one of which -- as we shall see -- for the sole purpose of constructing a Lyapunov operator is not essential. 

Of course, a most natural candidate for a Lyapunov operator is a time operator $T$ canonically conjugate to the Hamiltonian $H$, such that $T$ and $H$ form an imprimitivity system \cite{Mackey} (implying that each generates a translation on the spectrum of the other). Yet, a well known theorem of Pauli appears to tell us that this is impossible \cite{Pauli}. Recently, Galapon has attempted to bypass Pauli's arguments and find pairs of $T$ and $H$ satisfying the canonical commutation relations (CCR), but which are not in the class of imprimitivity systems \cite{Galapon}. Nevertheless, it can be shown that the $T$ operator that has been obtained in this way does not have the Lyapunov property. By contrast, other authors have given up on the conjugacy of $T$ and $H$.  In this context, Unruh and Wald's (UW) proof that a "monotonically perfect clock" does not exist \cite{Unruh} should be noted. However, while there is no widespread agreement on the definition of a time operator, most agree that it should have the Lyapunov property. In this paper it is our purpose to present a Lyapunov operator rather than a time operator, and it is precisely for this reason -- that beyond the Lyapunov property we do not impose any further requirements on our operator -- that the various no-go theorems previously mentioned do not apply to our construction.\\

We begin by presenting our arrow of time operator and proving that it has the required Lyapunov property. We do not attempt to derive or motivate its form, for which, see \cite{Yossi 07,Yossi 07 II}. Following this, we obtain its spectrum and eigenfunctions. Next, we address the two seemingly relevant no-go theorems by MPC and UW previously referred to, and show that they do not apply to our construction. We then go on to present the results of simulations illustrating the Lyapunov property for a large equivalence class of one dimensional potentials for
which M\o ller wave-operators exist. We end by discussing open questions and directions for future research.\\

Let us consider a Hamiltonian with finite uniform degeneracy and a continuous spectrum $E\in\left[0,\,\infty\right)$.
We claim that the operator 
\begin{widetext}
\begin{equation}
M_{F}=-\frac{1}{2\pi i}\sum_{j}\int_{0}^{\infty}dE\int_{0}^{\infty}dE'\frac{\left|E,\, j\right\rangle \left\langle E',\, j\right|}{E-E'+i0^{+}}\,,\label{arrow operator}\end{equation} where $E$ denotes the energy and  $j$ the degeneracy, is an ever decreasing Lyapunov operator \cite{foot1}. (Note the use of natural units $\hbar=c=1$.) To see this let us write down the expectation value of $M_{F}$ at time $t\geq 0$ with respect to some
arbitrary initial state $\left|\psi\right\rangle \in\mathcal{H}$
\begin{equation}
 \left\langle M_{F}\left(t\right)\right\rangle _{\psi}
=\left\langle \psi\left|e^{iHt}M_{F}e^{-iHt}\right|\psi\right\rangle 
= -\frac{1}{2\pi i}\sum_{j}\int_{0}^{\infty}dE\int_{0}^{\infty}dE'\frac{e^{i\left(E-E'\right)t}\psi_{j}^{\star}\left(E\right)\psi_{j}\left(E'\right)}{E-E'+i0^{+}}\,.\label{<M_F(t)>} \end{equation}
Using contour integration it is easy to verify that eq. (\ref{<M_F(t)>}) may be reexpressed as
 \begin{eqnarray}
\left\langle M_{F}\left(t\right)\right\rangle _{\psi} & = & -\frac{1}{4\pi^{2}}\sum_{j}\int_{0}^{\infty}dE\int_{0}^{\infty}dE'\int_{-\infty}^{\infty}d\sigma\frac{e^{i\left(E-\sigma\right)t}\psi_{j}^{\star}\left(E\right)\psi_{j}\left(E'\right)}{\left(E-\sigma+i0^{+}\right)\left(E'-\sigma-i0^{+}\right)}\nonumber \\
 & = & -\frac{1}{2\pi i}\sum_{j}\int_{0}^{\infty}dE\psi_{j}^{\star}\left(E\right)e^{iEt}\int_{-\infty}^{\infty}d\sigma\frac{e^{-i\sigma t}\tilde{f}_{j}\left(\sigma\right)}{E-\sigma+i0^{+}}\,,\label{<M_F(t)> II}\end{eqnarray}
with $\tilde{f}_{j}\left(\sigma\right)\hat{=}-\frac{1}{2\pi i}\int_{0}^{\infty}\frac{dE'\psi_{j}\left(E'\right)}{\sigma-E'+i0^{+}}$
(implying that $\tilde{f}_{j}\left(\sigma\right)\in\mathcal{L}^{2}\left(\mathbb{R}\right)$).
Then according to the Paley-Wiener theorem \cite{Paley-Wiener} $\tilde{f}_{j}\left(\sigma\right)$
is the Fourier transform of a function $f_{j}\left(\tau\right)\in\mathcal{L}^{2}\left(\mathbb{R}^{-}\right)$,
that is \begin{equation}
\tilde{f}_{j}\left(\sigma\right)=\int_{-\infty}^{0}d\tau e^{-i\sigma\tau}f_{j}\left(\tau\right)\,.\label{Paley-Wiener}\end{equation}
Substituting back into eq. (\ref{<M_F(t)> II}) we get
\begin{eqnarray}
\left\langle M_{F}\left(t\right)\right\rangle _{\psi} & = & -\frac{1}{2\pi i}\sum_{j}\int_{0}^{\infty}dE\psi_{j}^{\star}\left(E\right)e^{iEt}\int_{-\infty}^{\infty}d\sigma\frac{e^{-i\sigma t}}{E-\sigma+i0^{+}}\int_{-\infty}^{0}d\tau e^{-i\sigma\tau}f_{j}\left(\tau\right)\,,\nonumber \\
 & = & \sum_{j}\int_{0}^{\infty}dE\psi_{j}^{\star}\left(E\right)e^{iEt}\int_{-\infty}^{0}d\tau\Theta\left(-t-\tau\right)e^{-iE\left(t+\tau\right)}f_{j}\left(\tau\right)\nonumber \\
 & = & \sum_{j}\int_{-\infty}^{-t}d\tau\int_{0}^{\infty}dE\psi_{j}^{\star}\left(E\right)e^{-iE\tau}f_{j}\left(\tau\right)\,,\label{<M_F(t)> III}\end{eqnarray}
\end{widetext}where $\Theta\left(x\right)$ is the Heavyside function.
From the definition of $\tilde{f}_{j}\left(\sigma\right)$ and eq.
(\ref{Paley-Wiener}) we have that \begin{eqnarray}
f_{j}\left(\tau\right) & = & \frac{i}{4\pi^{2}}\int_{-\infty}^{\infty}d\sigma e^{i\sigma\tau}\int_{0}^{\infty}\frac{dE'\psi_{j}\left(E'\right)}{\sigma-E'+i0^{+}}\nonumber \\
 & = & \frac{1}{2\pi}\Theta\left(-\tau\right)\int_{0}^{\infty}dE'e^{iE'\tau}\psi_{j}\left(E'\right)\,.\label{f_k(tau)}\end{eqnarray}
Eq. (\ref{<M_F(t)> III}) now assumes the form \begin{equation}
\left\langle M_{F}\left(t\right)\right\rangle _{\psi}  =   2\pi\int_{-\infty}^{-t}d\tau\Big|\sum_{j}f_{j}\left(\tau\right)\Bigr|^{2}\,.\label{<M_F(t)> IV}\end{equation}
The expectation value of $M_{F}$ is thus seen to be nonnegative and
monotonically decreasing with time, irrespectively of the initial
state, tending to zero in the limit that $t$ goes to infinity.

To obtain the full spectrum of $M_{F}$, i.e. find its upper bound, we introduce the operator \begin{equation}
M_{B}=\frac{1}{2\pi i}\sum_{j}\int_{0}^{\infty}dE\int_{0}^{\infty}dE'\frac{\left|E,\, j\right\rangle \left\langle E',\, j\right|}{E-E'-i0^{+}}\,.\label{M_B}\end{equation} Similarly, $M_{B}$ can be shown to be an ever increasing
nonnegative operator. In particular, as $t$ tends to minus infinity the expectation value of $M_{B}$ tends to zero. Now \begin{equation}
M_{F}+M_{B}=\mathds{1}\,,\label{M_F+M_B}\end{equation}
and since $M_{B}$ is nonnegative as well, it follows that $M_{F}$ is bounded
from above by one.

The eigenstates of $M_{F}$ are found by solving the eigenvalue equation
$M_{F}\left|m,\, j\right\rangle =m\left|m,\, j\right\rangle $, $m\in\left[0,\,1\right]$,
with $j$ indicating the degeneracy. In the energy representation
the eigenvalue equation takes on the form \begin{equation}
-\frac{1}{2\pi i}\int_{0}^{\infty}dE'\frac{g_{m}^{\left(j\right)}\left(E'\right)}{E-E'+i0^{+}}=mg_{m}^{(j)}\left(E\right)\,.\label{eigenvalue equation}\end{equation}
Here $g_{m}^{\left(j\right)}\left(E\right)\hat{=}\left\langle E,\, j|m,\, j\right\rangle $;
the kernel's independence of $j$ allowing us to set $\left\langle E,\, j|m,\, k\neq j\right\rangle =0$.
We assume the existence of analytical functions $g_{m}^{\left(j\right)}\left(z\right)$,
which in the limit that $z\rightarrow E+i0^{+}$ equal $g_{m}^{\left(j\right)}\left(E\right)$.
This allows us to analytically continue eq. (\ref{eigenvalue equation})
into the complex plane \begin{equation}
-\frac{1}{2\pi i}\int_{0}^{\infty}dE'\frac{g_{m}^{\left(j\right)}\left(E'\right)}{z-E'}=mg_{m}^{(j)}\left(z\right)\,,\qquad\mathrm{Im}z\neq 0\,.\label{eigenvalue equation complex}\end{equation}
Taking the difference between the limits from above and below the
real axis we get \begin{eqnarray}
& & \int_{0}^{\infty}dE'\left(\frac{1}{E-E'+i0^{+}}-\frac{1}{E-E'-i0^{+}}\right)g_{m}^{\left(j\right)}\left(E'\right)\nonumber\\
& = & -2 \pi i m\left(g_{m}^{\left(j\right)}\left(E+i0^{+}\right)-g_{m}^{\left(j\right)}\left(E-i0^{+}\right)\right)\,,\label{limit difference}\end{eqnarray} and hence \begin{equation}
g_{m}^{\left(j\right)}\left(E\right)=m\left(g_{m}^{\left(j\right)}\left(E+i0^{+}\right)-g_{m}^{\left(j\right)}\left(E-i0^{+}\right)\right)\,.\label{limit difference II}\end{equation}
$g_{m}^{\left(j\right)}\left(z\right)$ can now be continued to a
second Riemann sheet by making use of the branch cut along $\left[0,\,\infty\right)$ in eq. (11).
The eigenvalue equation reduces to \begin{equation}
g_{m}^{\left(j\right)}\left(e^{2\pi i}z\right)=-\left(\frac{1-m}{m}\right)g_{m}^{\left(j\right)}\left(z\right)\,.\label{eigenvalue equation II}\end{equation}

The rotation on the left-hand side can be written using the dilation group defined via $D_{\alpha}f\left(z\right)=f\left(e^{\alpha}z\right)$. It is easy to check that the generator of this group is $z\frac{d}{dz}$, i.e. $e^{\alpha z\frac{d}{dz}}f\left(z\right)=f\left(e^{\alpha}z\right)$.
Taking $\alpha=2\pi i$, eq. (\ref{eigenvalue equation II}) can be
reexpressed as \begin{equation}
 e^{2\pi iz\frac{d}{dz}}g_{m}^{\left(j\right)}\left(z\right)
 =  -\left(\frac{1-m}{m}\right)g_{m}^{\left(j\right)}\left(z\right)\,,\label{eigenvalue equation III}\end{equation}
admitting solutions of the form $g_{m}^{\left(j\right)}\left(z\right)=N_{m}z^{\beta}$
with $\beta=\left(k+\frac{1}{2}\right)-\frac{i}{2\pi}\ln\left(\frac{1-m}{m}\right)$,
$k\in\mathbb{Z}$, and $N_{m}$ a normalization factor dependent on
$m$. Setting $k=-1$ and $N_{m}=\frac{1}{2\pi\sqrt{m\left(1-m\right)}}$
the solutions are orthogonal, i.e. satisfy $\int_{0}^{\infty}dEg_{m'}^{\left(j\right)\star}\left(E\right)g_{m}^{\left(j\right)}\left(E\right)=\delta\left(m-m'\right)$.
The full set of (delta-function normalized) eigenfunctions is therefore given by
\begin{equation}
g_{m}^{\left(j\right)}\left(E\right)=\frac{E^{-\frac{i}{2\pi}\ln\left(\frac{1-m}{m}\right)-\frac{1}{2}}}{2\pi\sqrt{m\left(1-m\right)}}\,.\label{eigenfunctions}\end{equation}

Let us now address MPC's and UW's no-go theorems. Without going into details, MPC claimed that within standard
quantum mechanics a nonequilibrium entropy operator cannot be defined. What is important for our purpose is that by a nonequilibrium entropy operator MPC mean a Lyapunov operator $\hat{\ell}$ such that $i\big{[}H,\,\hat{\ell}\big{]}=\mathcal{D}\geq0$. Their proof rests on the assumption that the measurement of $\hat{\ell}$ and its rate of change $\mathcal{D}$ should be mutually compatible,
i.e. $\big{[}\hat{\ell},\,\mathcal{D}\big{]}=0$. While one can debate whether this assumption is reasonable or not, 
it is easy to show that it is not satisfied by $M_{F}$. Hence, there is no conflict
with MPC's no-go theorem. 

UW proved that a monotonically perfect clock cannot be defined within the framework of standard quantum mechanics. 
By such a clock UW mean an ever increasing Lyapunov operator $\hat{\ell}$ with the additional property that it has a vanishing probability of {}``running backwards''. Thus, if we break up the spectrum of $\hat{\ell}$ into
an infinite succession of finite sized nonoverlapping intervals, and let $\left|\ell_{n}\right\rangle $ denote an eigenstate of the projection operator onto the $n$th interval centered about $\ell_{n}$, then for 
$t>0$ $\left\langle \ell_{m}<\ell_{n}\left|e^{-iHt}\right|\ell_{n}\right\rangle =0$. However, as is readily 
verified, $M_{F}$ does not share this extra property \cite{foot2}, and so no conflict arises with UW's no-go
theorem as well.\\

\begin{figure}[htp]
\centering
\includegraphics[scale=0.7]{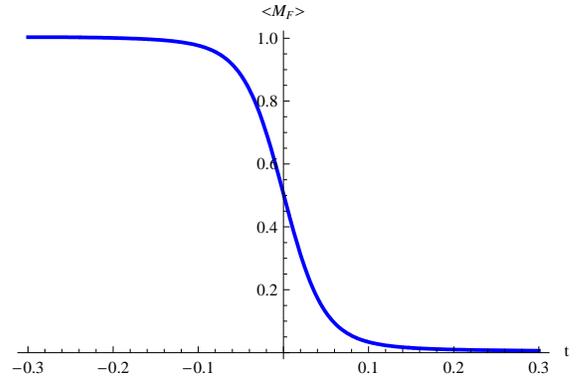}
\caption[]
{Monotonic decrease of $\left\langle M_{F}\right\rangle$. The figure depicts the monotonic decrease in time of $\left\langle M_{F}\right\rangle$ for a free Gaussian wave-packet with $p_{0}=6.4\mu$ and $\xi_{0}=3\mu$ describing the propagation of a free particle of mass $\mu$. $t$ is given in units of $[\mu]^{-1}$.}
\end{figure}

Next, we present the results of simulations illustrating the Lyapunov property. We consider the propagation from $x=-\infty$ to $x=\infty$ of a one dimensional free Gaussian wave-packet \begin{eqnarray}
 &  & \psi\left(x,\, t\right)\label{1D Gaussian wave-packet}\\
 & = & \left(\frac{\mu^{2}\xi_{0}^{2}}{\pi\left(\mu+i\xi_{0}^{2}t\right)^{2}}\right)^{\frac{1}{4}}\exp\left(-\frac{\mu \xi_{0}^{2}x^{2}+ip_{0}\left(p_{0}t-2\mu x\right)}{2\left(\mu+i\xi_{0}^{2}t\right)}\right)\,,\nonumber \end{eqnarray}
where $p_{0}$ and $\xi_{0}$ are the location and width of the wave-packet at $t=0$ in momentum space.
$M_{F}$ is given by \begin{equation}
M_{F}=-\frac{1}{i\pi \mu}\sum_{j=\pm}\int_{0}^{\infty}dpp\int_{0}^{\infty}dp'p'\frac{\left|p,\, j\right\rangle \left\langle p',\, j\right|}{p^{2}-p'^{2}+i0^{+}}\,,\label{arrow operator hydrogen}\end{equation}
with $\left|p,\,\pm\right\rangle $ denoting a plane wave state with a momentum $\pm p$, respectively, and $\mu$ the mass. Fig. (1) shows $\left\langle M_{F}\right\rangle $ as a function of time. If we now expand $\psi\left(x,\, t\right)$ in terms of the eigenfunctions of $M_{F}$, $\psi_{\pm}\left(m,\, t\right)=\left\langle m,\,\pm|\psi\left(t\right)\right\rangle $, then as time progresses the bulk of the state's support must shift from eigenfunctions having a higher value of $m$ to zero. See Fig. (2).

\begin{figure}[htp]
\centering
\includegraphics[scale=0.4]{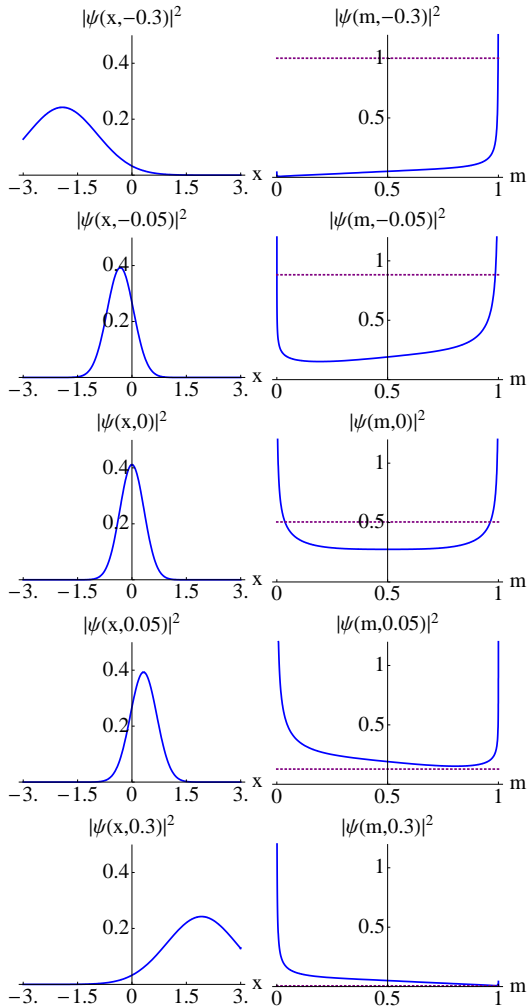}\caption[]
{Time frames of $\left| \psi\left(x,\,t\right)\right|^2$ and $\left| \psi\left(m,\,t\right)\right|^2$, $\psi\left(m,\,t\right) \hat{=} \left\langle m,\,+|\psi\left(t\right)\right\rangle+ \left\langle m,\,-|\psi\left(t\right)\right\rangle$ $t=-0.3,\,-0.05,\,0,\,0.05,\,0.3\,[\mu]^{-1}$, as computed for the Gaussian wave-packet of Fig. 1. $x$ is measured in units of $[\mu]^{-1}$. The dotted line gives the value of $\left\langle M_{F} \right\rangle$.
}\end{figure}

The behavior of the Lyapunov operator we have computed for the free particle evolution is precisely the same for a large equivalence class of Hamiltonians related to the free particle Hamiltonian, $H_0$, via the intertwining property of the M\o ller wave-operators $\Omega_{\pm}$. To see this let $H_{I}$ be any Hamiltonian of this class, then $ H_{I}=\Omega_{\pm} H_0 \Omega_{\pm}^{\dagger} $. In particular, $\left|E_{I} =E,\,\pm\right\rangle =\Omega_{+}\left|E_{0}=E,\,\pm\right\rangle $, implying that $M_{F}^{\left(I\right)}=\Omega_{+}M_{F}^{\left(0\right)}\Omega_{+}^{\dagger}$, with the zero index serving to denote the free system \cite{foot2.2}. If in the limit that 
$t\rightarrow-\infty$ $\left|\left\langle \psi_{0}\left(t\right)|\psi_{I}\left(t\right)\right\rangle \right|^{2}\rightarrow1$, i.e. both systems share the same asymptotic initial state,
then $\left|\psi_{I}\left(t\right)\right\rangle =\Omega_{+}\left|\psi_{0}\left(t\right)\right\rangle $. It follows that $\langle \psi_{I}\left(t\right)|M_{F}^{\left(I\right)}|\psi_{I}\left(t\right)\rangle = \langle \psi_{0}\left(t\right)|M_{F}^{\left(0\right)}|\psi_{0}\left(t\right)\rangle $ and $\langle m_I=m,\,\pm | \psi_{I}(t)\rangle=\langle m_0=m,\,\pm | \psi_{0}(t)\rangle$ \cite{foot2.5}.\\

To conclude, we have presented an arrow of time operator within the framework of standard quantum mechanics.  This operator can be defined for any system governed by a Hamiltonian with a uniformly finitely degenerate, absolutely continuous and semibounded spectrum \cite{foot3}. An immediate question that arises is whether our result can be generalized to Hamiltonians with different spectral properties. It is interesting that by discretizing $M_{F}-M_{B}=2M_{F}-\mathds{1}$ we obtain Galapon's $T$ operator \cite{Galapon}, which can be shown not to have the Lyapunov property, yet unlike our operator satisfies CCR  with the Hamiltonian.

In a forthcoming paper we will see how the existence of an arrow of time operator naturally leads to the existence of a new representation of the dynamics, in which the direction of time is manifestly exhibited. This "irreversible representation" is characterized by the property that the time evolution is a semigroup, and is particularly convenient for the description of processes such as the decay of unstable states, resonance processes, and other irreversible phenomena. 
\begin{acknowledgments}
Y. Strauss, J. Silman, and S. Machnes acknowledge the support of 
the Israeli Science Foundation (Grants no. 1282/05 and 784/06). Y. Strauss also acknowledges the support 
of the Center for Advanced Studies in Mathematics at Ben-Gurion University, where part of this research was conducted.\end{acknowledgments}


\begin{thebibliography}{10}
\bibitem{foot1}Whether $M_{F}$ is trivially generalized in the case
of infinite degeneracy, or whether its form is different, is the
subject of ongoing work.
\bibitem{foot2}To be more precise, $M_{F}$ does not share this property in that even though its expectation value is monotonically decreasing
in time, a single measurement may increase its expectation value.
\bibitem{foot2.2}Here it is implicitly implied that the wave-operators do not mix degenerate eigenstates. However, this assumption is not necessary, and the discrete indices on the two sides may just as well refer to different bases.
\bibitem{foot2.5}Note that this holds even if the spectrum of $H_I$ contains bound states, since the continuum maps only onto the continuum.
\bibitem{foot3}The generalization to the bounded spectrum case is straightforward, and will be presented in a forthcoming publication.
\end{thebibliography}
\end{document}